\documentclass{moriond}

\usepackage{cite}

\newcommand{\mumu}{\mu^+\mu^-}
\newcommand{\pipi}{\pi^+\pi^-}
\newcommand{\KK}{K^+K^-}
\newcommand{\mee}{e^+e^-}
\def\GeVM{~${\rm GeV}/c^2$}

\def\GeVp{~${\rm GeV}/c$}
\def\GeVE{~${\rm GeV}$}

\def\be{\begin{equation}}
\def\ee{\end{equation}}
\def\bea{\begin{eqnarray}}
\def\eea{\end{eqnarray}}

\begin{document}
\vspace*{4cm}
\title{MEASUREMENTS OF LOW ENERGY $\mee$ HADRONIC CROSS SECTIONS AND IMPLICATIONS FOR THE MUON g-2}

\author{ B. MALAESCU }

\address{Laboratoire de Physique Nucl\'eaire et des Hautes Energies, IN2P3-CNRS et Universit\'es Pierre-et-Marie-Curie et Denis-Diderot, 75252 Paris Cedex 05, France}

\maketitle\abstracts{
Numerous channels of the cross section $\mee \to {\rm hadrons}$ have been measured by the BABAR experiment using the ISR method.
For the $\pipi(\gamma)$ and $\KK(\gamma)$ channels, BABAR has pioneered the method based on the ratio between the hadronic mass spectra and $\mumu(\gamma)$.
Many systematic uncertainties cancel in the ratio, hence the precise measured cross sections.
These measurements have been exploited for phenomenological studies, like the determination of the hadronic contribution to the anomalous magnetic moment of the muon $(g-2)_\mu$.
}

\section{Introduction}

Precise measurements of the $\mee\to {\rm hadrons}$ cross-section are needed for various phenomenological studies, which motivated the BABAR extensive program for measuring them.
A well known example is the hadronic contribution to the muon magnetic moment anomaly~($a_\mu^{had}$).
It is dominated by the process $\mee\to\pipi(\gamma)$ which provides $73\%$ of the contribution, bringing also the dominant contribution to the uncertainty.
In these proceedings we present the $2\pi(\gamma)$ \cite{Aubert:2009ad_Lees:2012cj}, as well as the $2K(\gamma)$ \cite{Lees:2013gzt} precision measurements from BABAR.

\section{The BABAR ISR $\pipi$, $\KK$ and $\mumu$ analyses and the QED test} 

The measurements of the $\pi\pi$ and $\rm KK$ cross sections presented here use the ISR method~\cite{isr} for $\mee$ annihilation
events collected with the BABAR detector ($232~{\rm fb}^{-1}$ of data), at a center-of-mass energy $\sqrt{s}$ near $10.58$\GeVE.
We consider events $\mee \to X\gamma_{ISR}$, where $X$ can correspond to any final state, and the ISR~photon is emitted by the $e^+$ or $e^-$.
The $\pi\pi$, $KK$ and $\mu\mu$ spectra are measured.
These are the first NLO measurements, a possible additional radiation being taken into account in the analysis.
The $\mee \to \pi\pi(\gamma_{FSR})$ and $\mee \to \rm KK(\gamma_{FSR})$ cross sections are obtained as a function of the invariant mass of the final state $\sqrt{s'}$.
The advantage of the ISR method~(compared to an energy scan) is that all the mass spectrum is covered at once~(from threshold to $3~(5)$\GeVE~ for $\pi\pi~(\rm KK)$ in BABAR) with the same detector conditions and analysis.
The comparison between the measured muon spectrum and the NLO QED prediction is an important cross check of the analysis, called the QED test.
The cross section for the process $\mee\to X$ is related to the $\sqrt{s'}$ spectrum of $\mee\to X\gamma$ events through
$ {\mathrm{d}N_{X\gamma}} / {\mathrm{d}\sqrt{s'}}~=~\varepsilon_{X\gamma}(\sqrt{s'})~\sigma_{X}^0(\sqrt{s'})~ {\mathrm{d}L_{ISR}^{eff}} / {\mathrm{d}\sqrt{s'}}~, $
where $\sigma_X^0$ is the bare cross section~(excluding VP),
and $\varepsilon_{X\gamma}$ is the detection efficiency~(acceptance) determined by simulation with corrections obtained from data.
The $\pi\pi(\gamma_{FSR})$ and $\rm KK(\gamma_{FSR})$ cross sections are obtained from the ratio of the corresponding hadronic spectra and $\rm{L}^{\rm eff}_{\rm ISR}$ (derived using the muon spectrum).
The contribution of leading order FSR for muons is corrected for, while additional FSR photons are measured. 
The $\mee$ luminosity, additional ISR effects, vacuum polarization and ISR photon efficiency cancel in the ratio, hence the strong reduction of the systematic uncertainty. 
The selection of two-body ISR events is done requiring a photon with $E_\gamma^*>3$\GeVE~ and laboratory polar angle in the range $0.35-2.4~{\rm rad}$, 
as well as exactly two tracks of opposite charge, each with momentum $p>1$\GeVp~ and within the angular range $0.40-2.45~{\rm rad}$. 

The simulation of signal and background ISR processes is done with Monte Carlo (MC) event generators based on Ref.~\cite{eva}.
The structure function method is used to generate additional ISR photons, while {\small PHOTOS} is used for additional FSR photons~\cite{Aubert:2009ad_Lees:2012cj,Lees:2013gzt}.
MC- and, when possible, data-based studies are performed to evaluate the background level, found to be negligible for muons.
The simulation is used to compute the acceptance and mass-dependent efficiencies for trigger, reconstruction, PID, and event selection.
Specific studies are used to determine the ratios of data and MC efficiencies,
applied as mass-dependent corrections to the MC efficiency.
They amount to at most a few percent and are known to a few permil level or better.

Two kinematic fits to the $\mee\to X\gamma$ hypothesis (where $X$ allows for possible additional radiation) are performed for each event.
The two-constraint (2C) `ISR' fit allows an undetected photon collinear with the collision axis.
The 3C `FSR' fit is performed only when an additional photon is detected. 
Most events have small $\chi^2$ values for both fits.
An event with only a small $\chi^2_{ISR}$ ($\chi^2_{FSR}$) indicates the presence of additional ISR (FSR) radiation.
Events where both fits have large $\chi^2$ values result from multi-hadronic background, track or ISR photon resolution effects, or the presence of additional radiated photons.
To accommodate the expected background levels, different criteria in the ($\chi^2_{ISR}$,$\chi^2_{FSR}$) plane are applied.
The $\pi\pi$, $\rm KK$ and $\mu\mu$ masses are calculated from the corresponding best `ISR' or `FSR' fit.

The evaluations of the acceptance and $\chi^2$ selection efficiency are sensitive to the description of radiative effects in the generator. 
The difference of the FSR rate between data and MC is measured and results in a small correction for the cross section. 
Effects of the approximations in the simulation of additional ISR photons have been studied with the NLO {\small PHOKHARA} generator.
The differences occuring in acceptance yield corrections to the QED test. 
However, since radiation from the initial state is common to the pion, kaon and muon channels, the $\pi\pi(\gamma)$~($\rm KK(\gamma)$) cross section, obtained from the $\pi\pi$/$\mu\mu$~($\rm KK$/$\mu\mu$) ratio, are almost insensitive to the description of NLO effects in the generator.

\begin{figure}[tb]
  \centering
  \includegraphics[width=7.3cm]{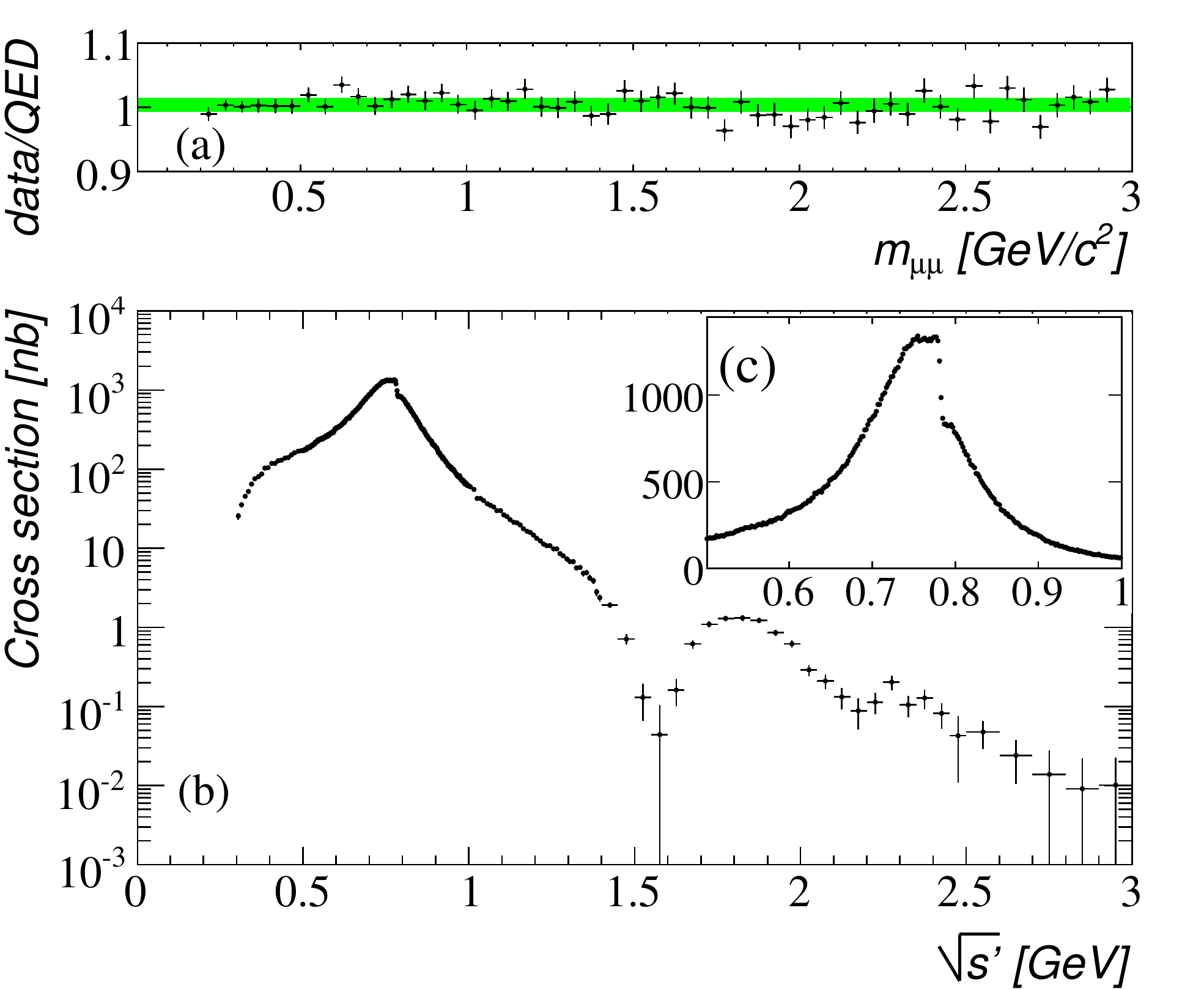}
  \includegraphics[width=8.5cm]{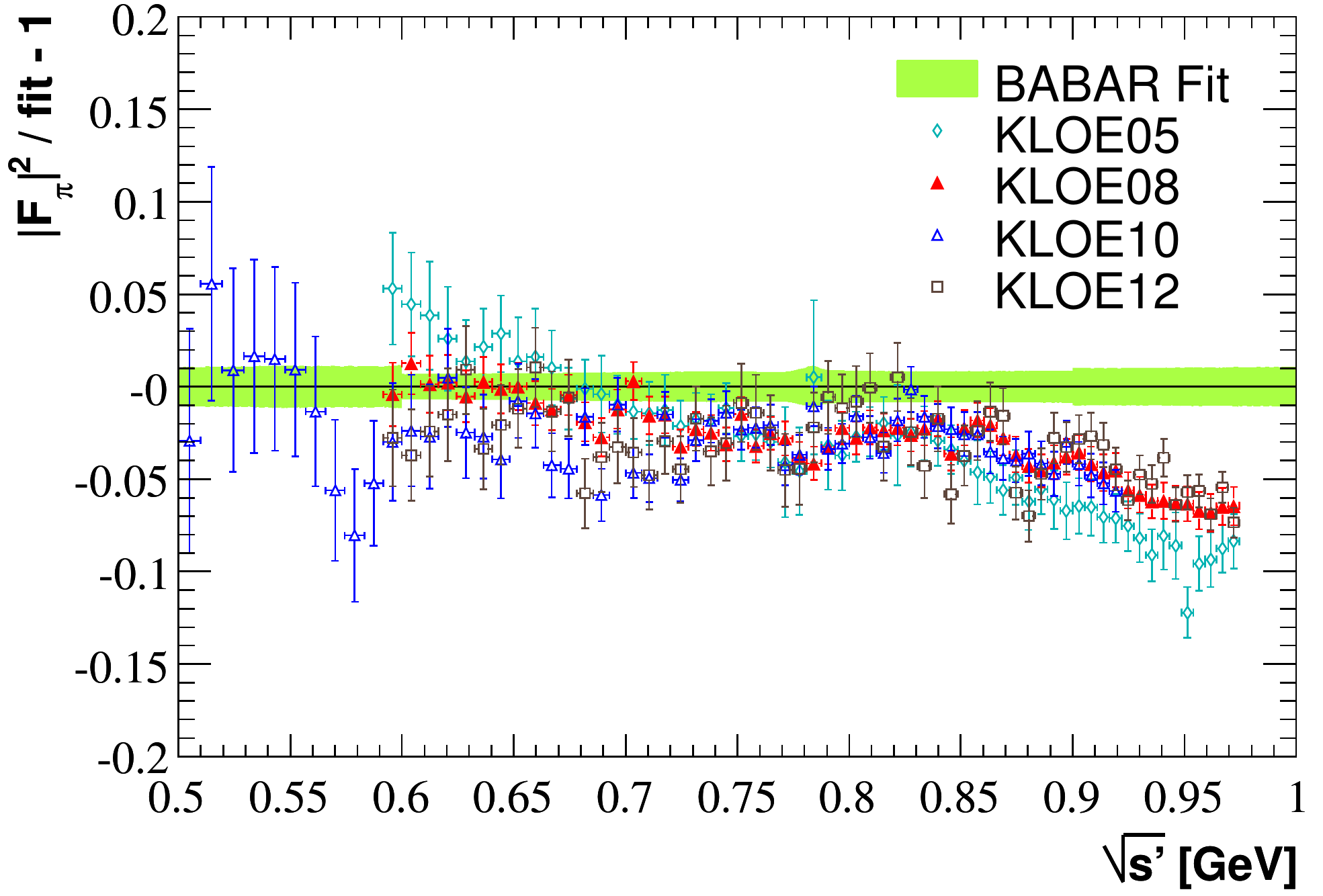}
  \caption{\small
Left:
(a) The ratio of the measured cross section for 
$e^+e^-\to\mu^+\mu^-\gamma(\gamma)$ to the NLO QED prediction. 
The band represents a fit to a constant value~(see text). 
(b) The measured bare cross section for $e^+e^-\to\pi^+\pi^-(\gamma)$ 
from 0.3 to $3$\GeVE. 
(c) Enlarged view of the $\rho$ region in energy intervals of 2 MeV.
Right:
Comparison between the $|F_\pi^2|$ from the various KLOE measurements and the BABAR fit.
}
\label{babar-log}
\end{figure}

The QED test involves two factors which cancel in the $\pi\pi$/$\mu\mu$~($\rm KK$/$\mu\mu$) ratio: $L_{ee}$ and the ISR photon efficiency, measured using a $\mu\mu\gamma$ sample selected only on the basis of the two muon tracks.
This test is expressed as the ratio of data to the simulated spectrum, after correcting for all known detector and reconstruction data-MC differences.
The generator is also corrected for its NLO deficiencies, using the comparison to {\small PHOKHARA}.
As shown in Fig.~\ref{babar-log}~(a), the ratio is consistent with unity from threshold to $3$\GeVM.
A fit to a constant value yields 
$ {\sigma_{\mu\mu\gamma(\gamma)}^{data}} / {\sigma_{\mu\mu\gamma(\gamma)}^{NLO~QED}} - 1 = (40\pm20\pm55\pm94)\times 10^{-4}~ $
($\chi^2/n_{\rm{df}}=55.4/54$),
where the uncertainties are statistical, systematic from this analysis, and 
systematic from $L_{ee}$ (measured using Bhabha scattering events), respectively.
The QED test is thus satisfied within a precision of 1.1\%.

\section{The $\pi\pi$ and $\rm KK$ cross sections and phenomenological applications}

A matrix-based unfolding of the background-subtracted $m_{\pi\pi}$~($m_{\rm KK}$) distribution~(corrected for data/MC efficiency differences) is performed to correct for resolution and FSR effects.
The precision of the unfolding procedure has been assessed using data-driven test models~\cite{bogdan}.

Fig.~\ref{babar-log} (b, c) shows the results for the $e^+e^-\to\pi^+\pi^-(\gamma)$ bare cross section including FSR, $\sigma^0_{\pi\pi(\gamma)}(\sqrt{s'})$.
The main features are the dominant $\rho$ resonance, the $\rho-\omega$ interference, a clear dip at $1.6$\GeVE~resulting from higher $\rho$ state interference, and some additional structure near $2.2$\GeVE.
A systematic uncertainty of only $0.5\%$ has been achieved in the central $\rho$ region. 
A VDM fit of the pion form factor~\cite{bogdanSlides} was exploited to compare the BABAR data to other experiments. 
The BABAR data are described well by this fit in the region of interest for the comparison. 
There is a relatively good agreement~(within uncertainties) when comparing to the Novosibirsk data~\cite{cmd-2,snd} in the $\rho$ mass region, while a slope is observed when comparing to the KLOE '08 data~\cite{kloe08}. 
A flatter shape is observed when comparing to the more recent KLOE~\cite{kloe10_12} data, obtained by the analysis of events with a detected, large angle ISR photon (see Fig.~\ref{babar-log} right). 
A good agreement is observed when comparing to the Novosibirsk and KLOE data, in the mass region below $0.5$\GeVM. 
There is a good agreement between the BABAR data and the most recent (isospin-breaking corrected)~$\tau$ data from Belle, while some systematic differences are observed when comparing to ALEPH and CLEO~\cite{bogdanSlides}.

The $\sigma_{\KK(\gamma)}^0(\sqrt{s'})$ cross section has been measured from the $\KK$ production threshold up to $5$\GeVE \cite{Lees:2013gzt}, and spans more than six orders of magnitude.
Close to threshold it is dominated by the $\phi$ resonance, while other structures are clearly visible at higher masses. 
The systematic uncertainty in the $\phi$ region is of only $0.7\%$.
We fit the kaon form factor with a model~\cite{FFK-kuehn} based on a sum of vector meson contributions, for measuring the $\phi$ resonance parameters~(found in good agreement with the world average) and providing an empirical parametrization of the form factor.
The measured charged kaon form factor is compared to data published by previous experiments~\cite{bogdanSlides}.
While the uncertainty of the BABAR cross section at the $\phi$ is $7.2\times 10^{-3}$, systematic normalization uncertainties of $2.2\%$ and $7.1\%$ are reported by CMD2 and SND, respectively.
The BABAR result, as well as the Novosibirsk measurements, are also affected by systematic uncertainties on mass calibration.
The observed mass differences are compatible with the BABAR and CMD2 (SND) calibration uncertainties.
However the normalization difference of $5\%$ is not consistent with the systematic uncertainties quoted by BABAR and CMD2.

The lowest-order contribution of the $\pi\pi(\gamma)$ intermediate state to the muon magnetic anomaly is given by a dispersion integral~\cite{kernel}.
The result of the integral from threshold to $1.8$\GeVE~, using the measured cross section and the full statistical and systematic covariance matrices, is
$ a_\mu^{\pi\pi(\gamma),LO} \:=\: (514.1 \pm 2.2 \pm 3.1)\times 10^{-10}~, $
where the uncertainties are statistical and systematic.
This value is larger than that from a combination of previous $e^+e^-$ data ($503.5\pm3.5$), but is in good agreement with the updated value from $\tau$ decays ($515.2\pm3.4$)~\cite{newtau}.
Using the $\pi^+\pi^-$ data from BABAR only, the deviation between the BNL measurement~\cite{bnl} and the theoretical prediction is reduced to $2.4 \sigma$.
The integral using the bare $e^+e^-\to\KK(\gamma)$ cross section from BABAR yields 
$a_\mu^{KK,\rm LO}\!=\!\left(22.93\pm0.18_{\rm stat}\pm0.22_{\rm syst}\pm0.03_{\rm VP}\right)\times10^{-10},$
for the energy interval between the $\KK$ production threshold and $1.8$\GeVE.
The first uncertainty is statistical, the second is the experimental systematic, while the third is from the $\phi$ parameters used in the VP correction.
This is the most precise result for the $\KK$ channel, and the only one covering the full energy range of interest.
For comparison, the combination of all previous data~\cite{Davier:2010nc} for the same range yields $\left(21.63 \pm 0.27_{\rm stat} \pm 0.68_{\rm syst}\right)\times10^{-10}$.

At large masses, the charged form factor can be compared to the asymptotic QCD prediction~\cite{chernyakFFK,BrodLepFFK}:
$ F_K(s) = 16\pi \,\alpha_s\left(s\right) \, f^2_{K^+} / s $ .
The fit of the squared form factor is performed between $2.5$ and $5$\GeVE~ with the function $A \alpha_s^2(s) /s^n$~($A$ and $n$ being free parameters),
which describes the data well ~($\chi^2/n_{\rm{df}}=23.4/32$).
It yields $n = 2.04 \pm0.22$, which is in good agreement with the QCD prediction $n=2$.
However, the fitted form factor is about a factor of 4 larger than the perturbative QCD prediction,
confirming the normalization disagreement observed with the CLEO measurements~\cite{CLEOK,Seth:2012nn}, at masses near the $\psi(2S)$ and above.

\section{Conclusions and perspectives} 

BABAR has analyzed the $\pi^+\pi^-$, $\KK$ and $\mu^+\mu^-$ ISR processes in a consistent way, from threshold to $3(5)$\GeVM. 
The absolute $\mu^+\mu^-$ cross section has been compared to the NLO QED prediction, the two being in agreement within $1.1\%$. 
The $e^+e^- \to \pi^+\pi^- (\gamma)$~($e^+e^- \to \KK (\gamma)$) cross section, derived through the ratio of the $\pi^+\pi^-$~($\KK$) and $\mu^+\mu^-$ spectra is rather insensitive to the detailed description of radiation in MC. 
A strong point of the present analysis comes from the fact that several uncertainties cancel in this ratio.

The BABAR data have been exploited for phenomenological studies, like fits and the computation of the hadronic contribution to $a_\mu$.
This contribution computed from the BABAR $\pi^+\pi^-$ spectrum, in the range $0.28-1.8$\GeVE, has a precision of $0.7\%$, similar to the precision of the combined previous measurements. 
For the contribution to $a_\mu$ from the $\KK$ channel, the BABAR result is almost three times more precise compared to the previous world average.

The BABAR $\pi^+\pi^- (\gamma)$ cross section data are in fair agreement with CMD2 and SND.
The agreement is poor when comparing with the various KLOE measurements, a dependence on the version of the KLOE measurements being observed too.
The comparison of the KLOE and $\tau$ data shows a discrepancy, while BABAR is in good agreement with the most recent $\tau$ results.

\end{document}